\documentclass[
               twocolumn,
               noshowpacs,          
               nopreprintnumbers,     
               aps,                 
               prd,                 
               a4paper,             
               superscriptaddress,  
               nofootinbib,         
               tightenlines,        
               floats,floatfix      
               ]{revtex4}

\usepackage{amsmath, amsfonts, amsthm, amssymb, graphicx}
\usepackage{epstopdf}
 \usepackage{slashed}
 \usepackage{graphicx,epsfig}
\usepackage{amsmath}
\usepackage {amssymb}
\usepackage[utf8]{inputenc}
\usepackage{hyperref}
\usepackage[francais]{babel}
\usepackage[T1]{fontenc}

\newcommand{\be}{\begin{eqnarray}}
\newcommand{\ee}{\end{eqnarray}}
\newcommand{\bea}{\begin{eqnarray}}
\newcommand{\eea}{\end{eqnarray}}

\begin{document}

\title{Deforming the Double Liouville String}

\author{Gaston Giribet}
\affiliation{Department of Physics, New York University, 726 Broadway, New York, NY10003, USA.}

\author{Mauricio Leston}
\affiliation{Instituto de Astronom\'{\i}a y F\'{\i}sica del Espacio IAFE-CONICET, Ciudad Universitaria, IAFE, 1428, Buenos Aires, Argentina.}
\affiliation{Departamento de F\'isica, Universidad de Buenos Aires FCEN-UBA and IFIBA-CONICET, Ciudad Universitaria, Pabell\'on 1, 1428, Buenos Aires, Argentina.}

\author{Pedro Schmied}
\affiliation{Departamento de F\'isica, Universidad de Buenos Aires FCEN-UBA and IFIBA-CONICET, Ciudad Universitaria, Pabell\'on 1, 1428, Buenos Aires, Argentina.}

\author{Bruno Sivilotti}
\affiliation{Departamento de F\'isica, Universidad de Buenos Aires FCEN-UBA and IFIBA-CONICET, Ciudad Universitaria, Pabell\'on 1, 1428, Buenos Aires, Argentina.}



\begin{abstract}
We consider a generalization of the double Liouville theory, which can be thought of as a two-parameter family of deformations of the so-called Virasoro Minimal String (VMS). The latter consists of a timelike ($c_-<1$ ) and a spacelike ($c_+>25$) Liouville field theory formulated on a fluctuating Riemann surface. For the deformed theory, we compute the sphere partition function exactly in $1/c_{\pm}$ and at third order in the coupling constant ($\lambda$) that controls the deformation. We also discuss the analogous computation in the case of the Complex Liouville String (CLS) theory, which is defined as two spacelike Liouville theories with complex central charges $c_{\pm }=13\pm i\mathbb{R}_{>0}$. We show that the partition functions of VMS and CLS differ at leading order in $\lambda $ due to the presence of elliptic functions in the observables of the latter. Both VMS and CLS theories have recently been studied in relation to many interesting models, including the double scaled Sachdev–Ye–Kitaev model, matrix models, and de Sitter gravity in 2 and 3 dimensions. We comment on the interpretation of the marginal deformation in some of these contexts.   

\end{abstract}

\maketitle

\section{Introduction} 
\label{sec:introduction}

In the last two years there has been a growing interest in conformal field theories composed of a pair of Liouville theories. The most prominent examples are the so-called Virasoro Minimal String (VMS) theory, proposed in \cite{Collier:2023cyw}, and the more recent Complex Liouville String (CLS) theory, studied in detail in \cite{Collier:2024kmo, Collier:2024kwt, Collier:2024lys,Collier:2024mlg}. These theories have recently been intensively investigated due to the close relationship they seem to have with other models  of interest for high energy physics; among these, certain matrix models, the double scaled Sachdev–Ye–Kitaev (SYK) model, and quantum gravity in de Sitter spaces in 2 and 3 dimensions; see \cite{Verlinde:2024zrh, Narovlansky:2023lfz, Goel:2023svz} and references thereof.

In this work, we will analyze both VMS and CLS in a broader framework, which we will call double Liouville string, which will allow us to treat both theories in a unified way. This demands a careful analysis of the analytical extension of the observables to be calculated. Here, we will consider a generalized double Liouville string theory, which can be thought of as a two-parameter family of deformations of VMS -- or CLS. 

The starting point is to consider a bosonic non-rational conformal field theory of the form
\begin{equation}
     \text{Liouville }_{c_-=1-\frac{\varepsilon}{2}}\  \oplus \  \text{Liouville }_{c_+=25+\frac{\varepsilon}{2}}\  \oplus  \ b\text{-}c\text{ ghosts}_{c=26}\nonumber
 \end{equation}
with total central charge $0$. The VMS is a theory of this type defined for $\varepsilon \in  \mathbb{R}_{+}$, with a timelike Liouville field with $c_-<1$ in direct sum with a spacelike Liouville field with $c_+>25$. Both Liouville fields have no direct coupling; however, they interact through the spacetime metric. The CLS is also a theory of this type, but defined for $\varepsilon \in -24 +i \mathbb{R}_{+}$, with two spacelike Liouville fields with $c_{\pm }\in 13\pm i\mathbb{R}_{+}$. In both cases, the critical condition $c_++c_-=26$ is fulfilled. 

For such theories we will consider a family of deformations controlled by two parameters. We will explicity compute the spherical partition function of the theory at third order in the coupling constant that controls the deformation. Our result will be exact in the (individual) central charge $c_{\pm }$.

In section II, we briefly review the main ingredients for our discussion, the spacelike and the timelike Liouville field theories. In section III, we introduce the double Liouville theory and explain the difference between the VML and the CLS theories. We will also introduce the marginal deformation of the theory. In Section IV, we compute the partition function for the deformed theory and obtain explicit expressions. We conclude in section V with a discussion about the possible interpretation of the marginal deformation in different setups. 

\section{Liouvile field theory}

\subsection{The spacelike Liouville theory}

Liouville field theory is defined by the action
\begin{align}
\label{Liouville_action}
    S_{\text{L}}[\phi] = \frac{1}{4\pi} \int_{\Sigma} d^2x\, \sqrt{g} \left(g^{\mu\nu} \partial_{\mu}\phi \partial_{\nu}\phi + Q R \phi + 4\pi \mu e^{2b\phi}\right) 
\end{align}
where $g_{\mu\nu}$ is the metric on the fluctuating sphere $\Sigma $, and $R$ is its scalar curvature. $Q$, $b$, and $\mu$ are three real parameters. The background charge $Q$ relates to $b$ as follows
\begin{align}
    Q = b+\frac{1}{b}.
\end{align}
The conformal field theory (CFT) defined in this way has central charge
\begin{equation}
c=1+6Q^2\geq 25\, .
\end{equation}

The primary states of the CFT are in correspondence with vertex operators of the form
\begin{equation}
V_{\alpha }(z)\, = \, e^{2\alpha \phi (z)}\label{vertices}
\end{equation}
which have conformal dimension $\Delta_{\alpha} =\alpha (Q-\alpha )$. The free parameter $\alpha$ can be regarded as the Liouville momentum. $\delta$-function normalizable states correspond to the values $\alpha \in \frac{Q}{2}+i \mathbb{R}$, \cite{Seiberg}.

Correlation functions of the vertex operators (\ref{vertices}) are defined as follows
\begin{equation}
\label{corr_vertex_1}
    \langle\prod_{i=1}^n V_{\alpha_i}(z_i)\rangle = \int_{\phi_{(\mathbb{CP}^1)}} 
    {D\phi\,\,  e^{-S_{\text{L}}[\phi]}}\,
    \prod_{i=1}^n e^{2\alpha_i\phi(z_i)} 
\end{equation}
where here we are defining the theory on the Riemann sphere. The 3-point function of the theory has been explicitly computed in \cite{Dorn:1994xn, Zamolodchikov:1995aa}; its form is known as the DOZZ formula, namely
\begin{align}
\label{3-point}
    \langle\prod_{i=1}^3 V_{\alpha_i}(z_i)\rangle  = \prod_{a<c}^3 |z_a-z_c|^{2\Delta-4(\Delta_a+\Delta_c)} C(\alpha_1, \alpha_2, \alpha_3)
\end{align}
where $\Delta_a=\alpha_a (Q-\alpha_a)$, $\Delta=\sum_{a=1}^3\Delta_a$, and where the structure constants reads
\begin{eqnarray}
    C(\alpha_1, \alpha_2, \alpha_3) &=& (\pi \mu \gamma(b^2) b^{2-2b^2})^{(Q-\alpha)/b}\times  \nonumber \\
    &&\frac{\Upsilon_b(b)}{\Upsilon_b(\alpha-Q)}\prod_{a=1}^3\frac{\Upsilon_b(2\alpha_a)}{\Upsilon_b(\alpha-2\alpha_a)}.\label{DOZZ}
\end{eqnarray}
where $\gamma(x) = \Gamma(x)/\Gamma(1-x)$, and $\alpha=\sum_{a=1}^3\alpha_a$. The special functions
\begin{eqnarray}
    \log \Upsilon_b(x) &=&\int_{\mathbb{R}_{+}} \frac{d\tau}{\tau} \left[ \left(x-\frac Q2\right)^2\, e^{-\tau}-\right.\nonumber \\
    &&\ \ \ \ \ \ \ \ \ \ \ \  \left. \frac{\sinh^2\left((\frac{Q}{2}-x)\frac{\tau}{2}\right)}{\sinh(\frac{b\tau}{2})\, \sinh(\frac{\tau}{2b})}\right]\label{representatione}
\end{eqnarray}
have been introduced in \cite{Zamolodchikov:1995aa}. Integral representation (\ref{representatione}) is defined on the interval $0<\text{Re}(x)<Q$, and the functions have to be analytically continued.

The $\Upsilon_b(x)$ functions obey interesting functional relations. Among the most relevant ones are the shift relations
\begin{align}
\label{ups_gamma_relation}
    \gamma(x\, b^{\pm 1})=&\frac{\Upsilon_b(x+b^{\pm 1})}{\Upsilon_b(x)}b^{\pm 2b^{\pm 1}x\mp 1}\, ,
\end{align}
and the reflection and duality relations
\begin{equation}\label{laticeeee}
\Upsilon_b(Q-x)=\Upsilon_b(x) \, , \ \ \ \ \ \Upsilon_b(x)=\Upsilon_{\frac 1b}(x)\, .
\end{equation}
The zeroes of $\Upsilon_b(x)$ are located at $x=-m\frac 1b-nb$ and $x=(m+1)\frac 1b +(n+1)b$ with $m,n \in \mathbb{Z}_{\geq 0}$, which defines a lattice invariant under (\ref{laticeeee}). Also, it holds $\Upsilon_b(Q/2)=1$.

The observables in Liouville field theory can be obtained in many different ways, including, of course, the 2D conformal bootstrap approach \cite{Zamolodchikov:1995aa}. A particularly useful method is the continuous approach, which amounts to combining the path integral method with the direct computation of correlators using the Coulomb gas formalism \cite{Goulian:1990qr}. This yields conformal integrals that can be solved in terms of $\Gamma$-functions as in the case of Minimal Models \cite{Dotsenko:1984ad}, and then analytically continued. In particular, in this way one can easily compute the partition function of the theory on the sphere topology. This reads \cite{Giribet:2022cvw}
\begin{align}
\label{Partition_function}
    Z_{\text{L}}[\mu] = \frac{(1-b^2)(\pi\mu\gamma(b^2))^{1+b^{-2}}}{\pi^3(b+b^{-1})\gamma(b^2)\gamma(b^{-2})}
\end{align}

This can be compared with the result obtained by a different method \cite{Zamolodchikov:2005jb}, which consists of solving the functional relation 
\begin{align}\label{dozz_partition}
    \frac{d^3Z_\text{L}[\mu]}{d\mu^3}    = -C(b,b,b).
\end{align}
Using the explicit form of the DOZZ structure constants (\ref{DOZZ}), one obtains
\begin{align}
     C(b, b, b) =& (\pi \mu \gamma(b^2) b^{2-2b^2})^{b^{-2}-2} \,\,\frac{\Upsilon_b^3(2b)}{\Upsilon_b(2b-b^{-1})\Upsilon_b^2(b)}\nonumber;
\end{align}
and, from this, using (\ref{ups_gamma_relation}) to express the $\Upsilon_b$-functions in terms of $\gamma$-functions, it is easy to verify that by integrating in $\mu$ one exactly reproduces the expression (\ref{Partition_function}).

There are other special features of formula (\ref{Partition_function}) that deserve attention. First, notice that it breaks self-duality under $b\leftrightarrow b^{-1}$. This has been observed by Zamolodchikov in \cite{Zamolodchikov:2005jb} and, in our opinion, requires further understanding. Notice also that (\ref{Partition_function}) does not agree with the $\alpha_i\to 0 $ limit of the Liouville 2-point function. More precisely, the 2-point function is
\begin{align}
\label{2-point}
    \langle V_{\alpha_1}(z_1)V_{\alpha_1}(z_2)\rangle  = |z_1-z_2|^{-4\Delta_{1}} \, B(\alpha_1)
\end{align}
with
\begin{align}
\label{2-point2}
    B(\alpha)=
(\pi \mu \gamma(b^2) )^{\frac{Q-2\alpha}{b}}\, \frac{\gamma(2\alpha b-b^2)\gamma(\frac{2\alpha}{b}-\frac{1}{b^2})}{\pi(2\alpha-Q)}
\end{align}
from what we observe that $B(0)=\frac{\pi^2}{(b^2-1)}Z_{\text{L}}[\mu ]$. This feature, which has also been observed in \cite{Zamolodchikov:2005jb}, can be explained by the operator fixing required to cancel the volume of the conformal Killing group. We will return to this later. 

\subsection{The timelike Liouville theory}\label{Liouville Timelike}

Now, let us move on to consider the timelike version of Liouville field theory, or timelike Liouville for short. It is defined by the action
\begin{align}
\label{Liouville_action_timelike}
    \hat{S}_{\text{L}}[\phi] = \frac{1}{4\pi} \int_{\Sigma} d^2x\sqrt{g} \left[-g^{\mu\nu} \partial_{\mu}\phi \partial_{\nu}\phi + \hat{Q} R \phi + 4\pi \mu e^{2b\phi}\right].
\end{align}
We use a circumflex on the quantities when they refer to the timelike case. Notice the {\it wrong} sign of the kinetic term in (\ref{Liouville_action_timelike}). Also, now we have
\begin{equation}
\hat{Q}=b-\frac 1b\, ,
\end{equation}
which yields the central charge
\begin{align}
    c = 1-6\hat{Q}^2 \leq 1\, .
\end{align}

While the action of the timelike theory (\ref{Liouville_action_timelike}) can actually be obtained from the spacelike action (\ref{Liouville_action}) by performing the Wick rotation\begin{align*}
    \phi \rightarrow -i\phi\, , \ \ \ \ 
    b \rightarrow ib\, ,
\end{align*}
the analytic extension of the theory is highly non-trivial, especially at the level of the observables \cite{Zamolodchikov:2005fy}. This can already be seen at the level of the 3-point function. In fact, in contrast to the naive expectation, the structure constants of the timelike Liouville theory are not simply obtained by replacing $b \to ib$ in DOZZ formula (\ref{DOZZ}). Remarkably, the result is rather {\it the inverse} -- in the sense that a normalization for the vertices exists such that the product of the timelike and spacelike structure constants is 1. For the timelike case one finds \cite{Zamolodchikov:2005fy, Harlow:2011ny, Giribet:2011zx}
\begin{eqnarray}
    \hat{C}(\alpha_1, \alpha_2, \alpha_3)
    &=&(\pi\mu\gamma(-b^2)b^{2+2b^2})^{(\hat{Q}-\alpha)/b}\times \label{timelike_DOZZ} \\
    &&\frac{\Upsilon_b(\hat{Q}+b-\alpha)}{b\Upsilon_b(b)}\prod_{a=1}^3\frac{\Upsilon_b(2\alpha_a-\alpha+b)}{\Upsilon_b(b-2\alpha_a)},\nonumber
\end{eqnarray}
which is actually quite different\footnote{Formula (\ref{timelike_DOZZ}) exhibits very special features which are not common in standard CFT. In particular, the limit $\hat{C}(0,\alpha_2,\alpha_3)$ does not vanish for $\alpha_2\neq \alpha_3$.} from (\ref{DOZZ}); see also \cite{Kostov:2005av, Schomerus:2003vv, Strominger:2003fn}. 

Expression (\ref{timelike_DOZZ}) can be derived by using the continuous approach \cite{Giribet:2011zx}. The proper analytic continuation of the conformal integral involved in the Coulomb gas formalism yields $\hat{C}(\alpha_1, \alpha_2, \alpha_3)\times \Gamma(0)$, where a divergent factor $\Gamma(0)$ is easily isolated, with the finite part exactly reproducing (\ref{timelike_DOZZ}). This divergence may likely be explained as due to a pole picked up when crossing a contour. In the path integral approach \cite{Harlow:2011ny} it is possible to clearly identify the integration cycles that lead to (\ref{timelike_DOZZ}). In the Coulomb gas approach, in contrast, this is less evident, and the prescription reduces to specify how one analytically continues the conformal integrals. In any case, it becomes evident that this issue of the divergent factor $\Gamma(0)$ is specific to the timelike theory since the analogous computation of the spacelike DOZZ formula (\ref{DOZZ}) does not yield such a factor. 

It is probably worth mentioning that the main trick to produce a formula such as (\ref{DOZZ}) and (\ref{timelike_DOZZ}) is to consider the functional relation 
\begin{equation}
\prod_{n=1}^{s}\gamma(nb^2)=\frac{\Upsilon_b(sb+b)}{\Upsilon_b(b)} \, b^{s((s+1)b^2-1)}\, ,
\end{equation}
which enables to write the product of quotients of $\Gamma$-functions in terms of $\Upsilon_b$-functions.

The reason why the timelike 3-point function does not agree with the naive analytic extension of the DOZZ formula to imaginary values of $b$ is related to the fact that DOZZ formula encounters an analytic boundary at $ib\in \mathbb{R}$, and so does the timelike expression \eqref{timelike_DOZZ}. This is summarized in the relation
\begin{eqnarray}
 C(i\alpha_1 , i\alpha_2 , i\alpha_3) &=& -i\,\hat{C}(\alpha_1 , \alpha_2 , \alpha_3)\frac{\vartheta_1'\left(0,\frac{1}{b^2}\right)}{\vartheta_1\left(-\frac{\alpha}{b} ,\frac{1}{b^2}\right)}\, \times \nonumber \\
 &&\ \ \ \ \ \prod_{a=1}^3 \frac{\vartheta_1\left(-\frac{2\alpha_a}{b} ,\frac{1}{b^2}\right)}{\vartheta_1\left(\frac{2\alpha_a-\alpha}{b} ,\frac{1}{b^2}\right)}
\end{eqnarray}
where $\vartheta (z,\tau)$ is the Jacobi eliptic function and where the prime stands for the derivative with respect to the first argument; see (\ref{bardelli}) below. Function $\vartheta (z,\tau)$ is well defined for $\text{Im}(\tau )>0$. As a consequence, the only solution for the spacelike structure constant, for which $b^2\in \mathbb{R}_+$, is (\ref{DOZZ}). Analogously, the only solution for the timelike structure constant, for which $-b^2\in \mathbb{R}_+$, is (\ref{timelike_DOZZ}). In contrast, for generic values $b^2\in \mathbb{C}$ two different solutions exist; see \cite{Zamolodchikov:2005fy} for a detailed explanation. This turns out to be very important for the calculation of the timelike partition function $\hat{Z}_{\text{L}}[\mu ]$. In fact, the timelike structure constant obeys
\begin{eqnarray}
    \hat{C}(b, b, b)=0\, ,
\end{eqnarray}
which turns out to be an obstruction to compute $\hat{Z}_{\text{L}}[\mu ]$ in the way we did it for the spacelike analog ${Z}_{\text{L}}[\mu ]$ in (\ref{dozz_partition}). Therefore, in the timelike theory we are forced to deal with the explicit computation in the continuous approach. This has been done in \cite{Giribet:2022cvw}; in order to avoid redundancies we refer to that paper; here, we will just sketch the main steps of the derivation: First, one integrates the zero-mode of $\phi $, which yields
\begin{eqnarray}
\label{part_func_timelike}
    \hat Z_{\text{L}} [\mu] &=& \frac{\mu^{\hat{m}+3}}{b} \Gamma(-\hat{m}-3)\int_{\mathbb{C}} \prod_{t=1}^{\hat{m}} d^2w_t \hspace{3pt}\prod_{t<l}^{\hat{m}}  |w_t - w_l|^{4b^2} \nonumber \\&& \times \, \prod_{r=1}^{\hat{m}} |w_r|^{4b^2} |1-w_r|^{4b^2} .
\end{eqnarray}
with $\hat{m} = \hat{s}-3$ and $\hat{s} = 1-\frac{1}{b^2}$. This produces an integrated $\hat{s}$-point free field correlator that can be computed by Wick contraction, using the free field propagator $ \langle\phi(z)\phi(w)\rangle =+ \log|z-w|$. Then, one performs the resulting integral expression assuming $\hat{m}\in \mathbb{Z}_{>0}$, and obtains
\begin{eqnarray}
    \hat{Z}_{\text{L}}[\mu] &=&\frac{\mu^{\hat{s}}}{b} \Gamma(-\hat{s})\Gamma(\hat{s}-2) \pi^{\hat{s}-3} \gamma^{\hat{s}-3}(1-b^2)\prod_{r=1}^{\hat{s}-3} \gamma(rb^2) \nonumber \\
    &&\prod_{t=1}^{\hat{s}-3}\gamma^2(1+(1+t)b^2)\, \gamma(-1-(\hat{s}-1+t)b^2).\nonumber
\end{eqnarray}
Finally, using properties of the $\gamma$-function one extends the result for $\hat{s}\in \mathbb{C}$. This yields
\begin{align}
\label{Partition_function_timelike}
    \hat Z_{\text{L}} [\mu] = \frac{(1+b^2)(\pi\mu\gamma(-b^2))^{1-b^{-2}}}{\pi^3(b-b^{-1})\gamma(-b^2)\gamma(-b^{-2})}.
\end{align}

Notice that this expression does agree with its spacelike counterpart (\ref{Partition_function}) after doing $b\to ib$. This explains why previous attempts to compute the partition function of the timelike theory by adapting the result of the spacelike theory in some cases may work. However, the lesson learned from the case of the 3-point function should prevent us from naively extrapolating the formula of the spacelike case. One finally obtains the relation
\begin{equation}\label{relevantlater}
Z_{\text{L}}[\mu_+]\hat{Z}_{\text{L}}[\mu_-]=\frac{(-1)^{b^{-2}}}{\pi^4b^2} \mu_-^{1-{b^{-2}}}\mu_+^{1+{b^{-2}}} \left(b^{2}\gamma(b^2)\right)^{\frac{2}{b^2}}\, .
\end{equation}
This relation will be relevant for us later. Still, it is worth not to mistake (\ref{relevantlater}) for the partition function of the VMS; see (\ref{esssssa}) below. Both quantities are closely related, but are not exactly the same. 

\subsection{Remark on the free field realization}

Regarding the calculation of the partition function and the structure constants using the Coulomb gas formalism, it is worth mentioning that in \cite{Zamolodchikov:1995aa} the calculation is carried out by introducing, in addition to the screening operator $e^{2b_{\pm}\phi }$, a second screening operator, which is often referred to as the dual cosmological operator that is obtained by transforming $b_{\pm}\to \pm 1/b_{\pm}$. In the free field calculation of the residues of the so-called resonant correlators, this second screening operator is necessary to determine the correct pole structure. Because of this, one might wonder whether it is not necessary to consider the dual operator in our analysis. It turns out that, thanks to the way in which we carry out the analytical continuation of the integral representation of the correlators like (\ref{part_func_timelike}), the insertion of a single screening operator suffices to reproduce the exact form of the correlators and, consequently, to reconstruct the full pole structure. This has been shown explicitly in \cite{Giribet:2001ft} for the non-compact WZW theory, and it works similarly in the case of Liouville theory. Another question that might arise is about the normalizability of the operators; for example, one might ask whether it is necessary to consider an analogue of the Seiberg bound for the operators that perform the deformation of the theory we are going to consider in the next section (see (\ref{operatore}) below). Again, the answer comes from the analytical continuation of the integral expressions, which leads to the exact form of the 3-point function for all values of the exponent $a_{\pm}$ of the exponential operator that realizes the perturbation.

\section{Double Liouville String}

\subsection{Double Liouville field theory}

Our formalism will allow us to treat the VMS and the CLS theories in a unified framework. As discussed above, this demands a careful treatment of the analytic extension of the observables. 

We start considering the conformal field theory that consists of two Liouville fields, namely
\begin{align}
\label{interacting_action}
    S[ \phi_+, \phi_-] = S_{\text{L}}[ \phi_+]+S_{\text{L}}[ \phi_-],
\end{align}
where $g_{\mu\nu}$ is the metric on the fluctuating sphere $\Sigma $. We can write each Liouville action as
\begin{eqnarray}
    S_{\text{L}}[\phi_{\pm}] &=& \frac{1}{4\pi} \int_{\Sigma}  d^2x\, \sqrt{g} \left( \,g^{\mu\nu} \partial_{\mu}\phi_{\pm} \partial_{\nu}\phi_{\pm} + Q_{\pm}R \phi_{\pm} \right.\nonumber \\
    &&\ \ \ \ \ \ \ \ \ \left. +\, 4\pi \mu_{\pm} e^{2b_{\pm}\phi_{\pm}}\,\right) \label{lambdacero1}
\end{eqnarray}
with the background charges $Q_{\pm} = b_{\pm}+\frac{1}{b_{\pm}}$, and the special condition $b_-=ib_+$ that guarantees that $c_-+c_+=26$. 

On the one hand, we have the CLS theory, which corresponds to $b_{+}=-ib_{-}=\beta e^{-i\frac{\pi }{4}}$ with $\beta \in \mathbb{R}$, with the two Liouville fields $\phi_{\pm}$ being spacelike -- in the sense that, for instance, their 3-point functions are given by the standard DOZZ formula (\ref{DOZZ}). This yields $c_{\pm }=13\pm i\, s$ with $s \in \mathbb{R}$. 

On the other hand, we have the VMS theory, which corresponds to the case $b_{+}=-ib_-\in \mathbb{R}$. This implies that we can treat the theory as if $\phi_-$ was a timelike Liouville field with real background charge $\hat{Q}_-=b_+-\frac{1}{b_+}$, while $\phi_+$ remains spacelike. 

The difference between VMS and CLS appears at the level of correlation functions, and manifests itself in the presence of Jacobi elliptic functions when combining the spacelike and the timelike DOZZ formulae accordingly. We will see below how this reflects in the form of the partition function of the marginally deformed theory.

\subsection{Deformation}

Now, let us introduce a marginal deformation. We define the theory
\begin{align}
\label{interacting_action}
   S_{\lambda}[\phi_+,  \phi_-]=S[ \phi_+,  \phi_-]+\lambda \int_{\Sigma} d^2x \sqrt{g}\, {O}_{a} ,
\end{align}
with the primary operator
\begin{align}
\label{operatore}
   {O}_{a}(z)=  e^{2a_+\phi_+(z) +2a_- \phi_-(z)} ,
\end{align}
where $\lambda$ is a new coupling constant and $a_{\pm }$ are two complex parameters that satisfy the marginality condition
\begin{align}
\label{parameters_interacting}
\begin{aligned}
     a_+(Q_+ -a_+)+a_-(Q_- - a_-)=1.
    \end{aligned}
\end{align}
Recall $Q_{\pm} = b_{\pm}+\frac{1}{b_{\pm}}$. This implies the relations $b_-=ib_+$ and $a_-=ib_+-ia_+$. The insertion of operator (\ref{operatore}) defines a two-parameter family of marginal deformations of the theory $S_0[\phi_+,\phi_-]=S[\phi_+,\phi_-]$. The independent parameters can be taken to be $a_+\in \mathbb{C}$ and $\lambda \in \mathbb{R}$. For $\lambda =0 $, one recovers the double Liouville theory (\ref{lambdacero1})-(\ref{parameters_interacting}). The same happens for $a_-=0$ or $a_+=0$, which corresponds to a simple shift of $\mu _+$ or $\mu_-$, respectively. The generic values of $a_{+}$ can be regarded as a perturbed Liouville field theory on the fluctuating sphere, where the spectator matter --using the terminology introduced in \cite{Zamolodchikov:2005jb}-- corresponds to a generalized minimal model with the insertion of two marginal operators.

\section{The partition function}

\subsection{Virasoro Minimal String}

We are interested in computing the spherical partition function of the marginally deformed double Liouville theory. By expanding the perturbation and integrating over the zero modes of the Liouville fields $\phi_{\pm}$, this quantities takes the form
\begin{align}\label{part_inter_general}
     &Z_{\text{VMS}}[\lambda ]=\frac{1}{b_+b_-}\sum_{k=0}^{\infty} \frac{(-\lambda)^k}{k!} \mu_+^{s_k^+}\mu_-^{s_k^-}\Gamma(-s_k^+)\Gamma(-s_k^-)\nonumber\\
     & \ \ \ \ \ \ \ \ \int_{\mathbb{C}}{\prod_{l=1}^kd^2u_l} \prod_{t=1}^{s_k^+} d^2w_t \prod_{r=1}^{s_k^-} d^2v_r
{\text{Vol}^{-1}{(PSL(2,\mathbb{C}))}}\nonumber\\ 
     &  \ \ \ \ \ \ \ \ 
     \Biggl\{\prod_{1\leq t<t'}^{s_k^+}|w_t-w_{t'}|^{-4b_+^2}\prod_{t=1}^{s_k^+}\prod_{l=1}^k|w_t-u_l|^{-4b_+a_+}\nonumber\\
     &\ \ \ \ \ \ \ \ \ \prod_{r=1}^{s_k^-}\prod_{l=1}^k|v_r-u_l|^{-4b_-a_-}\prod_{1\leq r<r'}^{s_k^-}|v_r-v_{r'}|^{-4b_-^2}\nonumber \\
     &\ \ \ \ \ \ \ \ \ \prod_{1\leq l<l'}^k|u_l-u_{l'}|^{-4(a_+^2+a_-^2)}\Biggr\},
\end{align}
where, for notational convenience, we have defined
\begin{equation}
s_k^+=1+\frac{1}{b^2_+}-k\frac{a_+}{b_+} \,  , \ \ \ \ \ s_k^-=2-k-s_k^+\, \label{relationote}
\end{equation}
for $k=0,1,2,3,..$. To produce this integral formula we have used exactly the same Coulomb gas techniques that shown to be powerful enough to solve both the spacelike and the timelike correlators. The $\Gamma (-s_{\pm})$ factors are generated when integrating over the zero modes of the fields $\phi_{\pm}$. $\text{Vol}(PSL(2,\mathbb{C}))$ is the volume of the conformal Killing group on the Riemann sphere, which has to be canceled to stabilize projective invariance. The cancellation can be accomplished by fixing three operators at, say, $z=0, 1, \infty$ in (\ref{part_inter_general}). We have many choices to implement this fixing. As a consistency check, we have explicitly verified that all such choices yield the same result. This is highly non trivial as it follows from the condition (\ref{parameters_interacting}) and the particular relation between $b_+$ and $b_-$. 

Another important feature of the integral expression (\ref{part_inter_general}) that one has to take into account in order to solve it is the relation (\ref{relationote}) between $s_k^+$ and $s_k^-$. This requires an analytic extension of the products involved. This is usual in the Coulomb gas realization of non-rational CFT and it amounts to extends the products 
\begin{equation}
\Pi(f|n)=\prod_{k=1}^{n}f(k) \ , \ \ \ n\in\mathbb{Z}_{>0}
\end{equation}
to the negative values of $n$, namely
\begin{equation}
\Pi(f|n)\equiv\prod_{k=0}^{-n-1}f^{-1}(-k)
 \ , \ \ \ n\in\mathbb{Z}_{<0}\, .
\end{equation}

Remarkably, while intimidating looking, the multiple integral in (\ref{part_inter_general}) can be solved explicitly -- at least up to third order in $\lambda$-- using the techniques we just explained. 

We define the expansion of the regularized partition function
\begin{equation}
\bar{Z}_{\text{VMS}}[\lambda] \,=\, \sum_{k=0}^{\infty}\, \lambda^{k}\, Z^{(k)}\label{laZeta}
\end{equation}
where the coefficients $Z^{(n)}$ depend on $\mu _+, \mu _- , b_+$ and $ a_+$. Recall that $b_-=ib_+$ and $a_-=ib_+-ia_+$. We have absorbed a $\Gamma (0)$ divergent factor appearing in the computation, analogously as in the case of the timelike 3-point function. The bar over $\bar{Z}_{\text{VMS}}[\lambda]$ in (\ref{laZeta}) is to make it explicit that we are referring to the renormalized partition function. This yields the finite part of each order $Z^{(n)}$. The explicit form of these coefficients are given below.

\subsubsection*{The $0^{\text{th}}$ orden}\label{orden_0_seccion}

The partition function at order $\mathcal{O}(\lambda^0)$ can be easily computed by integrating the first term of the infinite sum in (\ref{part_inter_general}), which yields a conformal integral of the type solved in \cite{Dotsenko:1984ad}. After reorganizing the result using properties of the $\Gamma $-function, such as $\gamma(b)\gamma(1-b)=1$ and $\gamma(1+b^2)=-b^4\gamma(b^2)$, one obtains
\begin{equation} \label{el0}
Z^{(0)}  =\frac{1}{\pi i} \frac{\mu_+^{s_0^+}}{\mu_-^{s_0^+-2}}\frac{(-1)^{s_0^++1}b_+^{4s_0^+-6}}{(1-b_+^4)}\gamma^{2s_0^+-2}(b_+^2)
\end{equation}
with $s_0^+=1+\frac{1}{b_+^2}$, $s_0^-=1-\frac{1}{b_+^2}$. Written in terms of $b_+$, this reads
\begin{equation}\label{esssssa}
Z^{(0)}=\frac{(-1)^{b_+^{-2}}}{i\pi b_+^2(1-b_+^4)} \mu_-^{1-{b_+^{-2}}}\mu_+^{1+{b_+^{-2}}} \left(b_+^{2}\gamma(b_+^2)\right)^{{2}{b_+^{-2}}}\, .
\end{equation}

It would be incorrect to expect this result to exactly match the product $Z_{\text{L}}[\mu_+]\hat{Z}_{\text{L}}[\mu_-]$ in (\ref{relevantlater}). This is similar to what happens between the 2-point function (\ref{2-point})-(\ref{2-point2}) and the partition function (\ref{dozz_partition}), which agree up to a simple factor; expression (\ref{el0}) equals $Z_{\text{L}}[\mu_+]\hat{Z}_{\text{L}}[\mu_-]$ up to a simple factor $\frac{i(1-b_+^4)}{\pi^3}$. The essential difference in the calculation of (\ref{esssssa}) and of (\ref{relevantlater}) is given by the volume of the conformal Killing group, which must be canceled jointly between the two Liouville pieces. In other words, the two Liouville theories interact via the metric. For example, the overall factor $i$ in (\ref{esssssa}) comes from the integration over the zero mode of each field $\phi_{\pm}$, which produces a factor ${b^{-1}_+}{b^{-1}_-}$. 

Expression (\ref{el0}) is exact in $b_+$, i.e. in $1/c_{\pm}$. Notice that, as expected, $|Z^{(0)}|$ is manifestly invariant under the change $\{\mu_{+}, b_+, a_+\}\leftrightarrow \{\mu_{-}, b_-, a_-\}$. The same is true for the higher orders.

\subsubsection*{The $1^{\text{st}}$ order}

The integral that appears at order $\mathcal{O}(\lambda^1)$ can also be solved in terms of $\Gamma$-functions. This yields
\begin{align}\label{1st_order}
    Z^{(1)} =\frac{1}{\pi i}\frac{\mu_+^{s_1^+}}{\mu_-^{s_1^+-1}}\frac{(-1)^{s_1^++1}b_+^{4s_1^+-5}}{(Q_+-a_+)(1-a_+b_+)}\frac{\gamma^{2s_1^+-1}(b_+^2)}{\gamma(1+b_+^2-2a_+b_+)}
\end{align}
with $s_1^+=1+\frac{1}{b_+^2}- \frac{a_+}{b_+}$, $s_1^-=-\frac{1}{b_+^2}+ \frac{a_+}{b_+}$. In addition to the poles at $a_+=Q_+ $ and $a_+=\frac{1}{b_+}$, $Z^{(1)}$ exhibits a series of simple poles at $a_+= \frac{b_+}{2}- \frac{n}{2b_+}$ with $n\in \mathbb{Z}_{\geq 0}$.

The fact that the order $\mathcal{O}(\lambda )$ contribution is not vanishing may seem puzzling, as at first sight it seems to correspond to the factorization of two 1-point functions of the operator (\ref{operatore}). However, it is not quite the case as both Liouville fields are defined on the same surface, and so there is a unique $\text{Vol}(PSL(2,\mathbb{C}))$ factor to cancel out by fixed insertions. Using the words of Al. Zamolodchikov, a conventional CFT wisdom requires the 1-point function [on the sphere] to vanish. \cite{Zamolodchikov:2005jb}. Poincar\'e invariance demands the 1-point function to be a constant; scale invariance demands the scaling dimension to be 0. In a conventional unitary CFT, where the only dimension-0 operator is the identity, this suffices to associate the non-vanishing 1-point function to the partition function. However, timelike Liouville is far from being a conventional CFT, in particular because it is believed to contain another operator of dimension 0 that does not obey the usual fusion rules \cite{Harlow:2011ny}. One can verify that, indeed, the fact that the order $\mathcal{O}(\lambda )$ is finite boils down to the existence of a non-trivial dimension-0 operator in the theory, which makes the limit $\hat{C}(0,0,a_-)\neq 0$. As we will see, this is different in the case of the CLS theory, where the order $\mathcal{O}(\lambda )$ does vanish.

\subsubsection*{Conformal points}

The fact that the other $\mathcal{O}(\lambda)$ is not zero might seem worrying since it might be interpreted as a symptom of the operator (\ref{operatore}) being marginal but not {\it exactly} marginal. Answering this question in timelike Liouville rigorously is difficult because the 3-point function $\hat{C}(b_-,b_-,a_-)$ is actually not well defined; it is divergent and, in the calculation of (\ref{1st_order}), appears multiplied with a vanishing factor coming from the spacelike Liouville piece. More precisely, in VMS we have that $\hat{C}(b_-,b_-,a_-)$ develops a double pole while $C(0,0,a_+)$ vanishes, yielding a finite product. The product $\hat{C}(0,0,a_-)C(b_+,b_+,a_+)$ is also finite. Recall the relation
\begin{equation}
 \langle \,e^{2b_{\pm}\phi_{\pm}} e^{2b_{\mp}\phi_{\mp}} e^{2(a_{+}\phi_{+}+a_{-}\phi_{-})}\, \rangle \,=\, - \frac{\partial^3 Z_{\text{VMS}}[\lambda ]}{\partial\lambda\partial\mu_{\pm}\partial\mu_{\mp}}
\end{equation}

The anomalous dimension of the operator that controls the deformation is given by $\hat{C}(a_-,a_-,a_-){C}(a_+,a_+,a_+)$. A non-vanishing value of this product would imply that the deformation is not exactly marginal. Therefore, one may wonder for what values of the parameters this last product of structure constants is zero. This happens for the values
\begin{equation}
a_+=\frac{m}{2b_+}+\frac{b_+}{2}\, , \ \ \ m\in \mathbb{Z}_{>0}\, \label{conformalpoints}
\end{equation}
which can be regarded as non-trivial zeroes. We will see later that these are precisely the values for which $\bar{Z}_{\text{VMS}}$ and $\bar{Z}_{\text{CLS}}$ do not receive corrections.

\subsubsection*{The $2^{\text{nd}}$ order}

Similarly, we can compute the order $\mathcal{O}(\lambda^2)$, which, after a lengthy manipulation, takes the following form
\begin{align}
\label{2nd_order}
    Z^{(2)}=&\frac{1}{2\pi i }\left(\frac{\mu_+}{\mu_-}\right)^{s_2^+}\frac{(-1)^{s_2^++1}b_+^{4s_2^+-3}}{Q_+-2a_+}\frac{\gamma^{2s_2^+}(b_+^2)}{\gamma^2(1+b_+^2-2a_+b_+)}
\end{align}
with $s_2^+=1+\frac{1}{b_+^2}- 2\frac{a_+}{b_+}$, $s_2^-=-1-\frac{1}{b_+^2}+ 2\frac{a_+}{b_+}$. This expression exhibits a series of double poles at $a_+= \frac{b_+}{2}- \frac{n}{2b_+}$ with $n\in \mathbb{Z}_{\geq 0}$.

\subsubsection*{The $3^{\text{rd}}$ order}

Remarkably, the order $\mathcal{O}(\lambda^3)$ can also be computed explicitly. Full-simplifying it, the final expression reads
\begin{align}
\label{3rd_order}
    Z^{(3)}=&\frac{1}{6 \pi i}\frac{\mu_+^{s_3^+}}{\mu_-^{s_3^++1}}\frac{(-1)^{s_3^++1}b_+^{4s_3^+}\gamma^{2s_3^++1}(b_+^2)}{\gamma^3(1+b_+^2-2a_+b_+)}.
\end{align}
with $s_3^+=1+\frac{1}{b_+^2}- 3\frac{a_+}{b_+}$, $s_3^-=-2-\frac{1}{b_+^2}+ 3 \frac{a_+}{b_+}$. This expression exhibits a series of triple poles at $a_+= \frac{b_+}{2}- \frac{n}{2b_+}$ with $n\in \mathbb{Z}_{\geq 0}$.

\subsubsection*{Higher orders}

Higher orders are difficult to compute as the conformal integrals become much more involved for $k\geq 4$. Nevertheless, an integral expression for all $k$ can be written down. The $k^{\text{th}}$ order in the expansion takes the generic form
\begin{equation}
Z^{(k)}= \frac{\gamma(b^2)^{k-2}}{i\pi \mu_-^{k-2}} \left(\frac{\mu_+}{\mu_-} b_+^4\gamma^2(b_+^2)\right)^{s_k^+} \gamma^k (2a_+b_+-b_+^2)\, z^{(k)}
\end{equation}
with $s_k^+=1+\frac{1}{b^2_+}-k\frac{a_+}{b_+}$ and $z^{(k)}$ being real coefficients independent of $\mu_{\pm }$. It is natural to conjecture that, for all $k\geq 3$, the poles and the zeroes of $Z^{(k)}$ will be determined by the factor $Z^{(k)}\propto \gamma^{k} (2a_+b_+-b_+^2)$. That is to say, the corrections $Z^{(k)}$ with $k>1$ vanish for $a_+=\frac{b_+}{2}+\frac{n}{2b_+}$ for $n\in \mathbb{Z}_{\geq 1}$, which are exactly the conformal points (\ref{conformalpoints}). In particular, this implies that the results reduces to $Z^{(0)}$ when the perturbation (\ref{operatore}) is given by $O_{\frac Q2}(z)=e^{Q_+\phi_+(z)+Q_-\phi_-(z)}$.

Taking $\mu_+\propto \pm \mu_-$, we observe that the order $\mathcal{O}(\lambda^k)$ of $Z_{\text{VMS}}[\lambda ]$ goes like $\sim \mathcal{O}(\mu^{2-k})$. This can be observed by the Knizhnik-Polyakov-Zamolodchikov (KPZ) scaling of the correlators. More precisely, if $\mu_-=\mu_+ \gamma(b_+^2)\gamma(1+b_+^2)$, then the exponent $s_k^+$ disappear from all $Z^{(k)}$.

The expressions for $Z^{(k)}$ ($k=0,1,2,3$) obtained above satisfy a series of non-trivial consistency checks: First, it turns out that, when computed with the Coulomb gas approach, all the terms in the $\lambda$ expansion exhibit the same overall divergence $\sim \Gamma (0)$, exactly as it happens in the computation of the timelike 3-point function. It is remarkable that this method suffices to efficiently isolate the divergent piece. That is to say, after normalizing $Z^{(0)}$ all the terms $Z^{(k)}$ result automatically finite. Second, one can verify that all combinations of insertions of different marginal operators to cancel the volume of the conformal Killing group yield the same result -- and this only happens after considering the precise relation between $a_-$ and $a_+$ that makes the deformation to be marginal. Third, for all the terms calculated above, it happens that, when $a_{\mp }=0$, so that $a_{\pm}=b_{\pm}$, the different orders in $k$ precisely organize as the Taylor expansion of the unperturbed solution with coupling constant $\mu_{\pm}\mp\lambda$; recall that those are precisely the values of $a_{\pm }$ for which the perturbation (\ref{operatore}) is expected to produces only a shift in the Liouville cosmological constant. 
%

\subsection{Complex Liouville String}

Now, let us perform a similar computation for the CLS theory. To do that, we have to take into account that CLS is defined in terms of two spacelike theories. Therefore, in contrast to the VMS, where seemingly magic cancellation between the $\Upsilon_b$-functions of the spacelike and the timelike theories take place, in the CLS we have to take care of the product of the function $\Upsilon_b$ and its extension $\Upsilon_{ib}$. In order to do that, it is convenient to consider the relation
\begin{eqnarray}
\label{ups_complex}
    \Upsilon_b(x)\Upsilon_{ib}(ib-ix)&=&
    e^{i\frac{\pi}{2}\left(x^2+\frac{x}{b}-xb+\frac{b^2}{4}-\frac{3}{4b^2}-\frac{1}{4}\right)}\nonumber \\&&\times \, \frac{\vartheta_1\left(\frac{x}{b}, \frac{1}{b^2}\right)}{\vartheta_1(\frac{1}{2}+\frac{1}{2b^2}, 1/b^2)},
\end{eqnarray}
where $\vartheta_1$ are the Jacobi elliptic functions, defined as 
\begin{eqnarray}
    \vartheta_1(z,\tau) = i \sum_{n=-\infty}^{\infty} (-1)^n e^{i\pi \tau(n-\frac 12)^2 + 2\pi iz(n-\frac 12)} \hspace{10pt} \label{bardelli}
\end{eqnarray}
for $\text{Im} (\tau ) >0$. These are entire functions with respect to $z$ for $\tau$ in the upper half $\mathbb{C}$ plane, and have simple zeros at $z=m+n\tau$ with $m,n\in\mathbb{Z}$. For this reason, as mentioned before, in order to use relation \eqref{ups_complex} we must require $1/b^2$ to have a positive imaginary part, i.e. $b = \beta e^{-i\theta /2}$ with $\beta \in \mathbb{R}$ and $\theta \in (0,\pi)$. The parameterization $b_{\pm} = \beta e^{\mp i\pi/4}$ does satisfy this requirement, cf. \cite{Verlinde:2024zrh}.

In addition, the function $\vartheta_1$ satisfies the following periodicity conditions
\begin{align}\label{periodicidad}
\begin{aligned}
    \vartheta_1(z+1, \tau) =& \, e^{-i\pi}\vartheta_1(z, \tau),\\
    \vartheta_1(z+\tau, \tau) =& \, e^{i\pi(1-\tau-2z)}\vartheta_1(z, \tau).
    \end{aligned}
\end{align}

When performing the calculation of the partition function of the CLS theory following the method used above, one finds the product of $\Upsilon_b$ functions coming from the two Liouville factors. This leads to the appearance of quotients of elliptic functions in some of the coefficients of the expansion in $\lambda $. In particular, one finds that
\begin{equation}
\begin{aligned}
\label{0,1,2,3_order_theta}
        Z^{(1)}\, \propto \,  \frac{\vartheta_1^3\left(0,\frac{1}{b_+^2}\right)\vartheta_1\left(\frac{2a_+}{b_+},\frac{1}{b_+^2}\right)}{\vartheta_1^3\left(\frac{a_+}{b_+},\frac{1}{b_+^2}\right)\vartheta_1\left(-\frac{a_+}{b_+},\frac{1}{b_+^2}\right)}
\end{aligned}
\end{equation}
and also
\begin{equation}
\begin{aligned}
\label{0,1,2,3_order_theta}
     Z^{(3)}\, \propto\, \frac{\vartheta_1\left(0,\frac{1}{b_+^2}\right)\vartheta_1^3\left(\frac{2a_+}{b_+},\frac{1}{b_+^2}\right)}{\vartheta_1\left(\frac{3a_+}{b_+},\frac{1}{b_+^2}\right)\vartheta_1^3\left(\frac{a_+}{b_+},\frac{1}{b_+^2}\right)}.
\end{aligned}
\end{equation}
These coefficients vanish. Hence, the final result reads
\begin{equation}
\bar{Z}_{\text{CLS}} [\lambda ] \, = \, Z^{(0)}+\lambda^2Z^{(2)}+\mathcal{O}(\lambda^4)
\end{equation} 
where $Z^{(0)}$ and $Z^{(2)}$ are given in (\ref{el0}) and (\ref{2nd_order}), respectively. This is the main result of this paper. Let us say that, strictly speaking, ours is an eduucated proposal for the result as it relies on the analytic extension of the Coulomb gas expressions, which are a priori valid only for a non-negative integer number of screening insertions.

\section{Discussion}

In this paper we have considered the double Liouville field theory deformed with a marginal operator controlled by a coupling constant $\lambda $ and labeled by an extra parameter $a_+$. For this theory, we explicitly computed the partition function on the fluctuating sphere at third order in $\lambda $. This provides a close form for a quantity of a theory that can be regarded as a generalization of the recently introduced VMS and CLS theories. 

As commented in the Introduction, the VMS and the CLS theories have recently been investigated in connection to many interesting models, including matrix models, the double scaled SYK model, and de Sitter gravity in 2 and 3 dimensions. Therefore, a natural question arises as to what the marginal deformation represents in these different setups. In the case of the matrix model associated to the VMS, the operator realizing the deformation can likely be treated in a standard way. In fact, such deformations -- or similar ones-- have been studied in detail in several matrix models. More interesting is the question of what the marginal deformation means in applications to de Sitter gravity. In order to consider how it might be possible to answer this question, let us review the relationship between the double Liouville theory and the Jackiw-Teitelboim (JT) 2-dimensional gravity: The non-interacting double Liouville theory can be rewritten as a sine-dilaton (or sinh-dilaton) theory \cite{Blommaert:2024ydx}. To show this, one first defines the action of the double Liouville theory, $S[\phi_-, \phi_+]$, on the locally flat metric $\delta_{\mu \nu }$. Then, one considers the complex field redefinition $
 {b_{\pm}\phi_{\pm} \equiv \rho \pm i\pi b_+^2 \Phi}$. After doing this, the two Liouville actions translate into the JT theory with a potential; namely
\begin{align}
    S_{\text{JT}} [g,\Phi ]=\frac{1}{2} \int d^2x \sqrt{g}\left(\Phi R + 4\, {\sin ({2\pi b_+^2 \Phi})} \right)\,,
\end{align}
where we have set $\mu_+ = - \mu_- $ to a specific value and where the metric is $g_{\mu \nu }=e^{2\rho} \delta_{\mu \nu }$. So, with this in mind, it is natural to ask what the marginal operator we turned on in the double Liouville theory corresponds to on the 2-dimensional gravity side. The same redefinition of fields gives us the answer; operator (\ref{operatore}) in (\ref{interacting_action}) takes the form
\begin{align}
   \delta S [g,\Phi ]=  2\lambda \int d^2x  \sqrt{g}\, e^{ 2\pi i   b_+(2{a_+}-b_+ )\Phi}\,,\label{perturbada}
\end{align}
which is real provided $\frac{a_+}{b_+}\in \frac 12 + i \mathbb{R}$. Notice that the exponent in (\ref{perturbada}) becomes an integer phase $e^{2\pi i n \Phi}$ with $n\in \mathbb{Z}_{> 0}$ precisely at the zeroes of $Z^{(k>0)}$; in particular, this includes the conformal points (\ref{conformalpoints}).  It would be interesting to investigate what the geometric implications of considering such a deformation in the gravitational theory are, and whether this could lead to anything interesting from the point of view of de Sitter gravity. In particular, it would be interesting to see whether the results obtained in this paper may have some applications to quantum gravity in de Sitter space. Another interesting question is what is the exact relation between these theories based on the analytic continuation of Liouville field theory and other non-local versions of Liouville theory that have been studied in connection to SYK \cite{nonlocal}. We leave these questions for future work.

\section*{Note added}

After the preprint of this work appeared on arXiv, the paper \cite{Collier:2025pbm} appeared, which deals with related topics. In particular, in that work a calculation of the partition function $Z_{\text{CLS}}$ on the sphere is presented, which turns out to be infinite. The authors of \cite{Collier:2025pbm} interpreted that result as being in tension with ours, which, strictly speaking, is incorrect since there is no obvious mismatch between both results. In fact, our result for $Z_{\text{CLS}}$ is infinite. Note that we never claimed that the Coulomb gas formalism allows to ``remove'' the divergence of $Z_{\text{CLS}}\sim\Gamma(0)$, but that said formalism allows to isolate it in a way that allows to give a proposal for its finite part $\bar{Z}_{\text{CLS}} $, in a similar way to what happens in the calculation of the 3-point function of timelike Liouville theory. A priori, there is no contradiction between the results of \cite{Collier:2025pbm} and our result. It would be interesting to see if, using the formalism of \cite{Collier:2025pbm}, there is a natural way to extract a finite result for the partition function. As observed there, the infinite result seems difficult to be interpreted from the point of view of gravity in dS space. As for our result for $\bar{Z}_{\text{VMS}}$, it is worth saying that the same (finite) result is obtained by integrating the product $C(b_+,b_+,b_+)\hat{C}(b_-,b_-,b_-)$ in $\mu_{\pm}$ following the trick (\ref{dozz_partition}). This is consistent with the the partition function of VMS being finite.

\[\]

\subsection*{Acknowledgments}
G.G. thanks Matthew Kleban, Beatrix M\"uhlmann, and Cameron Norton for discussions on related subjects. He also benefited from very interesting talks given at NYU by Victor Rodriguez and Herman Verlinde. It is a pleasure to thank the anonymous referee for his/her questions and remarks, which have helped clarify the discussion in this paper.

\[ \]

\providecommand{\href}[2]{#2}\begingroup\raggedright\endgroup
\end{document}